\documentclass[twocolumn,english,prl,aps,groupedaddress,raggedbottom,bibnotes,showpacs]{revtex4}
\usepackage[T1]{fontenc}
\usepackage[latin9]{inputenc}
\usepackage{color}
\usepackage{babel}
\usepackage{bm}
\usepackage{amsmath}
\usepackage{amssymb}
\usepackage{graphicx}
\usepackage[unicode=true,pdfusetitle,bookmarks=true,bookmarksnumbered=false,bookmarksopen=false,breaklinks=false,pdfborder={0 0 1},backref=false,colorlinks=false]{hyperref}

\newcommand{\RE}{\text{Re}}
\newcommand{\IM}{\text{Im}}

\makeatother

\begin{document}

\title{Magnetic instability of the orbital-selective Mott phase}

\author{Markus Greger}

\author{Michael Sekania}

\altaffiliation[Current address: ]{Theoretical Physik I, University of W\"urzburg, Am Hubland, 97074 W\"urzburg, Germany.}

\author{Marcus Kollar}

\affiliation{Theoretical Physics III, Center for Electronic Correlations and Magnetism,
Institute of Physics, University of Augsburg, 86135 Augsburg, Germany}

\date{\today}
\begin{abstract}
We characterize the low-energy physics of the two-orbital Hubbard
model in the orbital-selective Mott phase, in which one band is metallic
and the other insulating. Using dynamical mean-field theory with the
numerical renormalization group at zero temperature, we show that
this phase has a ferromagnetic instability for any nonzero Hund's
rule exchange interaction, which can be understood in terms of an
effective spin-1 Kondo Hamiltonian. The metallic band therefore behaves
as a singular Fermi liquid for which the self-energy has a logarithmic
singularity at the Fermi energy. 
\end{abstract}

\pacs{71.27.+a, 71.30.+h}

\maketitle
In order to explain the properties of most strongly correlated metals,
their multi-orbital band structure must be taken into account. In
such strongly interacting multi-band systems the local Coulomb interaction
acts not only on the electronic charge, as in single-band systems,
but also on the spin degrees of freedom. The additional spin-spin
interaction is due to the Hund's rule exchange coupling and can lead
to a whole array of new complexity which is absent in the single-band
case. This relevance of the Hund's rule coupling for the physics of
many strongly correlated metals has been discussed on many occasions
in recent years~\cite{doi:10.1146/annurev-conmatphys-020911-125045,medici2011janus,Haule_njp}.
This development has led to the classification of so-called \emph{Hund's
metals}~\cite{Haule_njp}, i.e., itinerant systems in which the Hund's
rule coupling is primarily responsible for strongly correlated behavior.

Iron-based superconductors provide an important example for such systems:
On the one hand, they are increasingly considered to be strongly correlated
because they often exhibit typical correlated behavior such as small
coherence scales and significant mass enhancements~\cite{doi:10.1146/annurev-conmatphys-020911-125045}.
On the other hand their density-density interaction is only moderate,
so that traditionally they would be regarded as weakly correlated
because they are not close to a Mott metal-to-insulator transition
(MIT). The MIT itself can be already understood from the paradigmatic
single-band Hubbard model, which however does not contain the spin-spin
interaction and is therefore unable to describe Hund's metals. Besides
modifying the character of strong correlations~\cite{PhysRevLett.108.087004},
the complexity introduced by the Hund's rule coupling can also induce
rich metallic physics beyond the traditional Fermi-liquid (FL) picture
of the single-band Hubbard model in the paramagnetic phase, as well
as new quantum phase transitions in addition the Mott MIT.

As we will explicitly show in this paper, the much-studied orbital-selective
Mott phase (OSMP) represents precisely such a new genuinely multi-orbital
phase for which the Hund's rule coupling plays a crucial role, and
which we characterize in this paper (for temperature $T=0$). 
The concept of the OSMP was first introduced to explain the metallic
properties of Ca$_{2-x}$Sr$_{x}$RuO$_{4}$~\cite{anisimov2002orbital}
and describes cases in which certain bands are Mott insulators while
the other bands remain metallic and need not be close to localization.
This phase thus provides an interesting chimera between metal and Mott
insulator, which has been the subject of many theoretical
studies~\cite{PhysRevLett.102.126401,PhysRevLett.92.216402,PhysRevB.72.205124,PhysRevLett.95.116402,PhysRevB.87.205135,PhysRevB.72.201102}.
An OSMP was also identified in several other materials. Namely the
manganite compound La$_{1-x}$Sr$_{x}$MnO$_{3}$ with its localized
$t_{2g}$ and metallic $e_{g}$ electrons is regarded as one of the
prototypical realizations of the
phase~\cite{doi:10.1146/annurev-conmatphys-020911-125045}.  Other
examples are FeO and CoO under
pressure~\cite{PhysRevB.82.195101,PhysRevB.85.245110} as well as
V$_{2}$O$_{3}$~\cite{PhysRevB.73.045109,PhysRevB.76.085127}.  Of
particular interest is the recent observation of OSMPs in iron
pnictides such as A$_{x}$Fe$_{2-y}$Se$_{2}$ (A=K,
Rb)~\cite{PhysRevLett.110.067003,2013arXiv1309.6084} and
FeSe$_{0.42}$Te$_{0.58}$~\cite{PhysRevB.82.140508}, suggesting the
relevance of orbital-selective physics in microscopic models for the
pnictides~\cite{PhysRevLett.110.067003,2012arXiv1212.3966D,PhysRevB.87.045122}.
Significant departures from FL in the OSMP behavior were already
established in Ref.~\onlinecite{Biermann_PhysRevLett.95.206401} and
explained by mapping the lattice Hamiltonian onto an effective
double-exchange model at low energies. This effective model led to the
conjecture of an instability towards ferromagnetism
\cite{Biermann_PhysRevLett.95.206401}, competing with an
antiferromagnetic instability \cite{Koyama20093267} due to
superexchange between localized spin degrees of freedom.
Here we concentrate on the ground-state properties of the OSMP,
classify its non-FL nature, and show
that the Hund's rule exchange indeed causes a ferromagnetic
instability as soon as the OSMP is reached.

\emph{Two-band Hubbard model.---}
For our systematic study of the low-energy physics of the OSMP we use
its fundamental theoretical model~\cite{Biermann_PhysRevLett.95.206401,PhysRevB.72.205124},
i.e., the two-band Hubbard model with different bandwidths, on-site
Hubbard, density-density, and Hund's rule exchange interactions without
interorbital hopping, solved in dynamical mean-field theory (DMFT)~\cite{PhysRevLett.62.324,RevModPhys.68.13}.
The Hamiltonian is given by 
\begin{align}
H & =-\sum_{\langle ij\rangle m\sigma}t_{m}d_{im\sigma}^{\dagger}d_{jm\sigma}^{\phantom{\dagger}}+U\sum_{im}n_{im\uparrow}n_{im\downarrow}+H_{J},\nonumber \\
H_{J} & =\sum_{i\sigma\sigma'}(U_{1}-\delta_{\sigma\sigma'}J)n_{i1\sigma}n_{i2\sigma'}\label{eq:LattHam}\\
 & \;\phantom{=}\;+\tfrac{1}{2}J\sum_{im\sigma}d_{im\sigma}^{\dagger}(d_{i\bar{m}\bar{\sigma}}^{\dagger}d_{im\bar{\sigma}}^{\phantom{\dagger}}+d_{im\bar{\sigma}}^{\dagger}d_{i\bar{m}\bar{\sigma}}^{\phantom{\dagger}})d_{i\bar{m}\sigma}^{\phantom{\dagger}}.\nonumber 
\end{align}
Here $i,j$ denote site, $m$ $=$ $1,2$ orbital, and $\sigma$ $=$
$\uparrow,\downarrow$ spin indices, with bars denoting the respective
alternate value. We assume semi-elliptic densities of states with
bandwidths $W_{m}$, corresponding, e.g., to scaled nearest-neighbor
hopping amplitudes $t_{m}$ $=$ $W_{m}/(4\sqrt{{\cal Z}})$ on a Bethe
lattice in the limit of infinite coordination number ${\cal
  Z}$~\cite{RevModPhys.68.13}.  We consider half-filled bands and put
$W_{2}/W_{1}$ $=$ $2$ throughout the paper; other fillings and
bandwidth ratios yield qualitatively similar behavior. Note that the
Hamiltonian consists of two Hubbard models with the same Hubbard
interaction $U$ which are coupled at each site $i$ only by $H_{J}$
through the interorbital repulsion $U_{1}$ and Hund's rule exchange
coupling $J$. As a consequence, the single-particle Green functions
$G_{m}(\omega)$ and self-energies $\Sigma_{m}(\omega)$ are
band-diagonal, whereas the two-particle and higher-order Green
functions also have nondiagonal components.  Two-particle Green
functions will be important for the characterization of ground-state
properties in the OSMP and the question of its
stability~\cite{Biermann_PhysRevLett.95.206401,PhysRevB.72.205124}.
In the following we treat the charge interaction $U_{1}$ and the spin
interaction $J$ as independent parameters because they act in
different channels. In most cases we will put $U_{1}$ $=$ $U-2J$,
valid for $d$ electrons. Only homogeneous phases are considered.
Previously we studied the FL phase of this model (for which $U$ is so
small that both bands are metallic and no phase transition
occurs)~\cite{PhysRevLett.110.046403}, showing that a small Hund's
rule coupling induces the same coherence scale in both bands even
though their single-particle spectra are quite different.  We use the
numerical renormalization group (NRG)~\cite{RevModPhys.80.395} to
solve the effective DMFT impurity problem, making exponentially small
excitation energies accessible, employing the same code and parameters
as in Ref.~\onlinecite{PhysRevLett.110.046403}. Here we concentrate on
the low-energy physics and two-particle quantities at zero
temperature, a temperature regime that is notoriously difficult to
reach with Quantum Monte Carlo methods but required to study the
stability of the phase. In particular we fully characterize the
non-FL~\cite{Biermann_PhysRevLett.95.206401} properties of the
itinerant band.

\begin{figure}[tb]
  \begin{centering}
    \includegraphics[width=0.9\hsize]{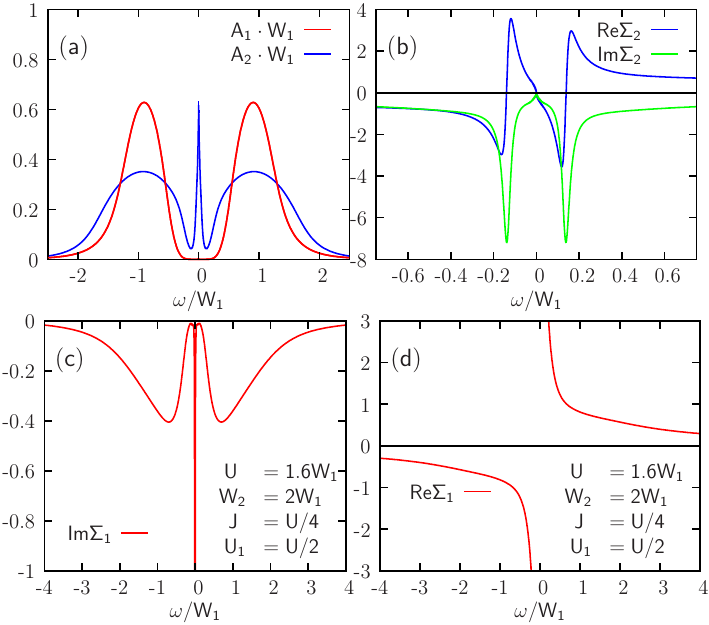}%
  \end{centering}
  \caption{\label{fig:FIG2}Single-particle quantities in the OSMP, i.e., (a)
    spectral functions $A_{m}(\omega)$ $=$ $-(1/\pi)\IM\, G_{m}(\omega)$
    and (b,c,d) self-energies $\Sigma_{m}(\omega)$, are qualitatively
    similar to the single-band Hubbard model in the metallic resp. insulating
    phase. Note that for the metallic (Fermi-liquid) band, $A_{2}(0)$
    is pinned to its non-interacting value $4/(\pi W_{2})$ in DMFT~\cite{RevModPhys.68.13}.}
\end{figure}

\begin{figure}[b]
  \begin{centering}
    \includegraphics[width=0.9\hsize]{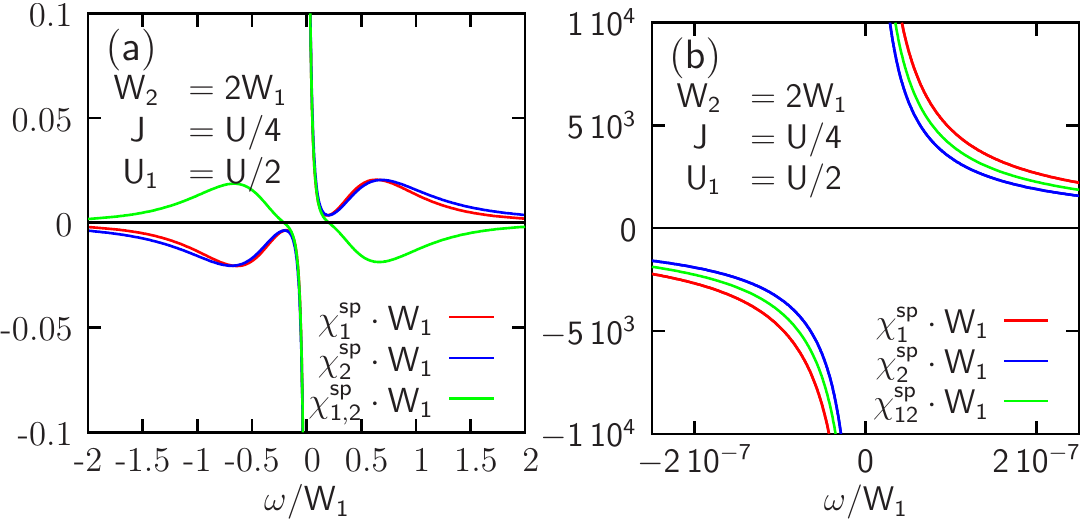}%
  \end{centering}
  \caption{\label{fig:FIG3}Spin susceptibilities in the OSMP (same
    parameters as Fig.~\ref{fig:FIG2}) for large (a) and vanishing
    frequencies (b). All three susceptibilities diverge for
    $\omega\rightarrow0$.  The low-frequency cutoff for the spectra in
    this calculation is $\omega_{c}\simeq10^{-8}W_{1}$.}
\end{figure}

In the OSMP the Hubbard interaction $U$ is sufficiently strong that
the narrow band ($m$ $=$ $1$) becomes Mott insulating, but
sufficiently weak that the wide band ($m$ $=$ $2$) remains
metallic. The obtained one-particle Green functions $G_{m}(\omega)$
and self-energies $\Sigma_{m}(\omega)$ are very similar to those of
two independent single-band Hubbard models~\cite{PhysRevLett.83.136},
one Mott insulating and one metallic, as shown Fig.~\ref{fig:FIG2}.
However, the two bands are not in fact independent, due to the interband
coupling terms in the Hamiltonian. This is evident only from two-particle
response functions, namely the spin susceptibilities $\chi_{m,m'}^{\text{sp}}(\omega)$
$=$ $-\IM\,\langle\langle S_{i,m}^{z},S_{i,m}^{z}\rangle\rangle_{\omega}/\pi$
($\chi_{m}^{\text{sp}}$ $\equiv$ $\chi_{m,m}^{\text{sp}}$ denotes
the diagonal parts), which exhibit a dominant low-frequency response,
see Fig.~\ref{fig:FIG3}(a). 
In the single-band Hubbard model this behavior is neither found in
the Mott insulating nor in the metallic phase and hence represents
a true multi-band effect induced by the interband coupling. In view
of the increasing frequency resolution of NRG for $\omega$ $\rightarrow$
$0$, the low-frequency closeup in Fig.~\ref{fig:FIG3}(b) provides
evidence that $\chi_{m}^{\text{sp}}$ indeed diverges in the OSMP,
suggesting a magnetic instability of the phase at zero temperature.

\begin{figure}[b]
  \begin{centering}
    \includegraphics[width=0.9\hsize]{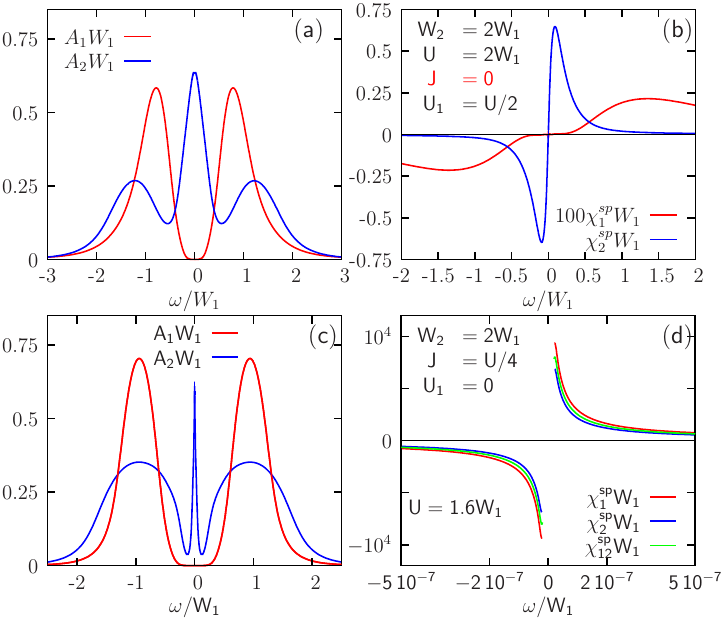}%
  \end{centering}
  \caption{\label{fig:FIG4}Comparison of the cases $J=0$, $U_{1}\neq0$
    (a, b) and $J\neq0$, $U_{1}=0$ (c, d).}
\end{figure}

This striking correlated behavior arises from the Hund's rule
exchange interaction $J$ rather than the density interaction $U_{1}$,
as can be seen from Fig.~\ref{fig:FIG4} 
in which either $J$ or $U_{1}$ is set to zero. For $J$ $=$ $0$,
$U_{1}$ $>$ $0$ the spin susceptibilities are essentially the same
as for two decoupled Hubbard models (one Mott insulating and one metallic),
i.e., the peaks in $\chi_{2}^{\text{sp}}(\omega)$ are finite while
$\chi_{1}^{\text{sp}}(\omega)$ has a spin gap. Note that the gaps
in $A_{1}(\omega)$ and $\chi_{1}^{\text{sp}}(\omega)$ imply that
the Mott-localized spins of the insulating band are decoupled from
the rest of the system at small excitation energies. Furthermore the
off-diagonal susceptibility $\chi_{12}^{\text{sp}}(\omega)$ vanishes
exactly for $J$ $=$ $0$, because its equation of motion contains
$[S_{z,1},H_{J}]$ $=$ $-[S_{z,2},H_{J}]$ $=$ $J(S_{1}^{+}S_{2}^{-}-S_{2}^{+}S_{1}^{-})$.
Hence the two bands are essentially decoupled both at the one- and
two-particle level. By contrast, for $U_{1}$ $=$ $0$ and arbitrarily
small $J$ $>$ $0$ the susceptibilities remain divergent and the
system cannot be regarded as the composition of one Mott-insulating
and one metallic single-band Hubbard model. Thus the OSMP has a quantum
critical point at $J$ $=$ $0$, with $U_{1}$ merely modifying its
properties quantitatively. We therefore set $U_{1}$ to zero in the
following and discuss nonzero $U_{1}$ again at the end. 

\begin{figure}[tb]
  \begin{centering}
    \includegraphics[width=0.9\hsize]{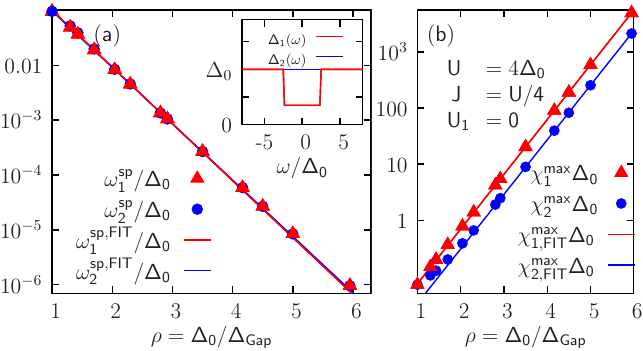}%
  \end{centering}
  \caption{\label{fig:FIG5} (a) Exponential decay of the quasiparticle
    coherence scale $\omega_{i}^{\text{sp}}/\Delta_{0}$ $=$ $A_{i}$
    $\exp(-B_{i}\rho)$ as a function of the pseudogap strength $\rho$
    $\equiv$ $\Delta_{0}/\Delta_{\text{gap}}$ ($A_{1}$ $=$ $0.1030$,
    $B_{1}$ $=$ $2.3626$, $A_{2}$ $=$ $0.1095$, $B_{2}$ $=$
    $2.3844$). (b) Amplitude $\chi_{i}^{\text{max}}$ of the spin
    susceptibilities. The exponential dependence
    $\chi_{i}^{\text{max}}\Delta_{0}$ $=$ $C_{i}$ $\exp(D_{i}\rho)$
    for $\rho\gtrsim2$ follows from the exponential dependence of the
    coherence scale (Kondo temperature) $\omega_{1}^{\text{sp}}$
    $\approx$ $\omega_{2}^{\text{sp}}$ which is the only scale that
    determines $\chi_{i}^{\text{sp}}(\omega)$ for small $\omega$ in
    the limit $\omega_{1}^{\text{sp}}\rightarrow0$ ($C_{1}$ $=$
    $0.0785$, $D_{1}$ $=$ $2.2441$, $C_{2}$ $=$ $0.0353$, $D_{2}$ $=$
    $2.2364$). The inset in (a) shows the hybridization
    functions~\eqref{eq:Deltagap}. Parameters for both plots given in
    (b).}
\end{figure}

\emph{Minimal two-impurity Anderson model.---}
In order to understand the divergent susceptibilities in Figs.~\ref{fig:FIG3}
and \ref{fig:FIG4}(d), we construct a minimal low-energy model that
captures the low-energy physics of the spin degrees of freedom in
the OSMP. As described below, such divergences are found in the two-impurity
Anderson model (TIAM)~\cite{PhysRevLett.110.046403,medici2011janus}
in which one impurity spin is localized (unhybridized with the host)
and the other itinerant (hybridized), in analogy to the OSMP. Its
Hamiltonian is given by~\cite{Koga20051366,PhysRevB.73.155106} 
\begin{align}
H_{\text{TIAM}} & =\sum_{\bm{k}m\sigma}\epsilon_{\bm{k}m}c_{\bm{k}m\sigma}^{\dag}c_{\bm{k}m\sigma}^{\phantom{\dagger}}+\sum_{m\sigma}\epsilon_{m}n_{m\sigma}\nonumber \\
 & ~~~+\sum_{\bm{k}m\sigma}\big(V_{\bm{k}m}c_{\bm{k}m\sigma}^{\dag}d_{m\sigma}^{\phantom{\dagger}}+\text{h.c.}\big)+H_{J}^{\text{loc}}\,,\label{eq:TIAM}
\end{align}
 where the local interaction $H_{J}^{\text{loc}}$ has the same form
as $H_{J}$, but without the index $i$. This is also the type of
TIAM onto which the Hamiltonian~\eqref{eq:LattHam} is mapped in DMFT, subject
to two self-consistency conditions for $G_{m}(\omega)$. The coupling
of the two impurity sites to the baths is characterized by the hybridization
functions $\Delta_{m}(\omega)$ $=$ $\sum_{\bm{k}}|V_{\bm{k}m}|^{2}/(\omega+i0-\epsilon_{\bm{k}m})$.
We consider a TIAM in which the hybridization function for the itinerant
band $\Delta_{2}(\omega)$ is constant, while for the Mott-insulating
band the hybridization function $\Delta_{1}(\omega)$ has a (pseudo-)gap,
i.e., a piecewise constant hybridization function which is smaller
in a low-$\omega$ range, \begin{subequations}\label{eq:Deltagap}
\begin{align}
\Delta_{1}(\omega) & =\Delta_{0}-(\Delta_{0}-\Delta_{\text{gap}})\Theta(\omega_{0}-|\omega|),\label{eq:Delta1gap}\\
\Delta_{2}(\omega) & =\Delta_{0},\label{eq:Delta2gap}
\end{align}
 \end{subequations}where $\Theta(x)$ is the unit step function and
we choose $\omega_{0}=2.5\Delta_{0}$ for the gap interval (see inset
in Fig.~\ref{fig:FIG5}a). (We verified that other, qualitatively
similar choices for the TIAM yield comparable results.) This model
interpolates between a standard TIAM ($\Delta_{\text{gap}}=0$) and
a TIAM with a fully gapped band ($\Delta_{\text{gap}}=\Delta_{0}$)
and mimics the self-consistent hybridization functions obtained from
DMFT.

\begin{figure}[b]
  \begin{centering}
    \includegraphics[width=0.99\hsize]{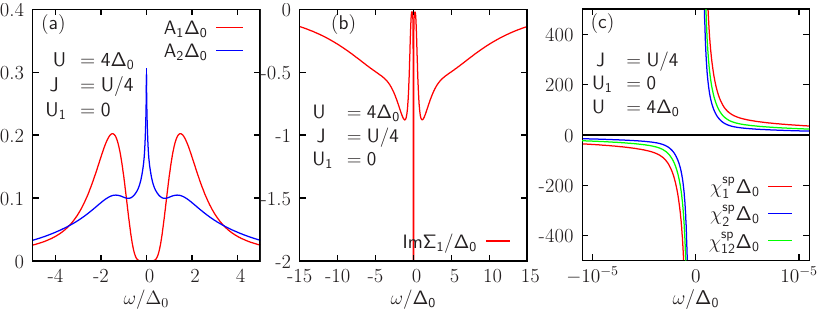}%
  \end{centering}
  \caption{\label{fig:FIG6}Spectral functions $A_{m}(\omega)$ (a),
    self-energy $\IM\,\Sigma_{1}(\omega)$ of the Mott-insulating band
    (b) and spin susceptibilities $\chi_{mm'}^{\text{sp}}(\omega)$ (c)
    for the fully gapped TIAM, i.e., for $\rho$ $=$ $\infty$. The
    quantities capture all qualitative features of the self-consistent
    DMFT results in the OSMP, shown Figs.~\ref{fig:FIG2} and
    \ref{fig:FIG3}. }
\end{figure}

The low-energy behavior of the spin susceptibilities are characterized
by the energy scale of spin fluctuations in the TIAM, i.e., the extrema
$\omega_{m}^{\text{sp}}$ of $\chi_{m}^{\text{sp}}(\omega)$ and the
peak amplitudes $\chi_{m}^{\text{max}}$ $\equiv$ $\chi_{m}^{\text{sp}}(\omega_{m}^{\text{sp}})$
\cite{PhysRevLett.110.046403}. These are shown in Fig.~\ref{fig:FIG5}
 as functions of $\rho$ $\equiv$ $\Delta_{0}/\Delta_{\text{gap}}$,
a parameter that measures how small the hybridization is inside the
gap interval. Both characteristic quantities exhibit an exponential
dependence on $\rho$. Note that the coherence scales of \emph{both}
bands vanish exponentially and are approximately equal, $\omega_{1}^{\text{sp}}$
$\approx$ $\omega_{2}^{\text{sp}}$, although the gap is opened \emph{in
only one} of the two hybridization functions. This correlation is
reminiscent of the Hund's-rule-induced proportionality of the two
self-energies in the FL phase of the model~\cite{PhysRevLett.110.046403}.
Furthermore, the corresponding peak amplitudes $\chi_{m}^{\text{max}}$
increase exponentially as the gap is opened. We thus conclude that
for a fully gapped TIAM ($\rho$ $=$ $\infty$) the low-energy scale
$\omega_{m}^{\text{sp}}$ is zero while the spin susceptibilities
$\chi_{m}^{\text{sp}}$ diverge. In Fig.~\ref{fig:FIG6} 
 we plot $\IM\,\Sigma_{1}(\omega)$, $A_{m}(\omega)$, and $\chi_{m,m'}^{\text{sp}}(\omega)$
for this fully gapped case. We observe a striking resemblance to the
corresponding DMFT results in the OSMP (Figs.~\ref{fig:FIG2}a,c
and \ref{fig:FIG4}b). Again we verified that other choices for the
interactions $J$ and $U_{1}$ lead to the same qualitative behavior
(not shown). The only important prerequisites for the divergences
in $\chi_{m,m'}^{\text{sp}}(\omega)$ are the low-energy gap in $\Delta_{1}(\omega)$
and a nonzero Hund's rule coupling $J$ $>$ $0$.

\emph{Effective two-impurity Kondo model and singular Fermi liquid.---}
The behavior of $\chi_{mm'}^{\text{sp}}(\omega)$ can be
understood from the Kondo limit ($U$ $\gg$ $\Delta_{0}$) of the
TIAM, i.e.,~\cite{medici2011janus,PhysRevLett.110.046403} 
\begin{align}
H_{\text{2IK}}=\sum_{\bm{k}m\sigma}\epsilon_{\bm{k},m}n_{\bm{k}m\sigma}+\sum_{m}J_{m}\bm{s}_{m}\cdot\bm{S}_{m}-J\bm{S}_{1}\cdot\bm{S}_{2}.\!\!\label{eq:KondoModel}
\end{align}
 Here $\bm{S}_{m}$ describes the spins of the two impurity orbitals
and $\bm{s}_{m}$ are the spins of the host electrons (with momentum
distributions $n_{\bm{k}m\sigma}$) at the impurity site. The superexchange
coupling $J_{m}$ $>$ $0$ between $\bm{S}_{m}$ and $\bm{s}_{m}$
is antiferromagnetic; the Hund's rule coupling $J$ $>$ 0 (from~\eqref{eq:LattHam})
provides the coupling $-J\bm{S}_{1}\cdot\bm{S}_{2}$ of the impurity
spins.

In the TIAM, the dependence of $\chi_{m}^{\text{sp}}(\omega)$ on
$\rho$ at low energies is due to spin fluctuations which are described
by~\eqref{eq:KondoModel}: when $\Delta_{1}(\omega)$ is fully gapped
($\rho$ $\rightarrow$ $\infty$), the antiferromagnetic coupling
$J_{1}\propto\Delta_{\text{gap}}$ between the spin $\bm{S}_{1}$
and its host band vanishes. The ferromagnetic coupling of the two
impurity spins will then produce a spin-$1$ object for any nonzero
$J>0$, i.e., it will favor the triplet sector of $\bm{S}=\bm{S}_{1}+\bm{S}_{2}$.
This composite spin-$1$ is coupled to the electrons from bath $\Delta_{2}(\omega)$
but decoupled from bath $\Delta_{1}(\omega)$, i.e., it is only partially
screened. In the self-consistent DMFT solution of the lattice model~\eqref{eq:LattHam}
a similar situation occurs in the OSMP: the gap in the self-consistent
bath $\Delta_{1}(\omega)$ implies that the spins of the gapped band
$A_{1}(\omega)$ are coupled to the rest of the system only through
$J$, leading to triplet formation across orbitals $1$ and $2$.

\begin{figure}[b!]
  \begin{centering}
    \includegraphics[width=0.9\hsize]{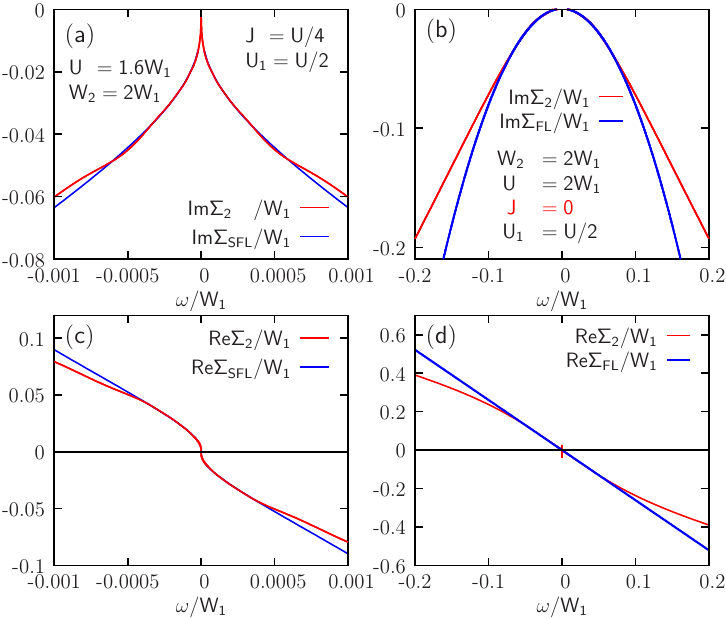}%
  \end{centering}
  \caption{\label{fig:FIG7}Low-energy behavior of the metallic
    self-energy $\IM\,\Sigma_{2}(\omega)$ ($\RE\,\Sigma_{2}(\omega)$)
    in the OSMP (a, c) and the corresponding fits to the SFL
    expressions Eqs.~\eqref{eq:imSIGMASFL} and \eqref{eq:reSIGMASFL};
    fit parameters are $a_{1}$ $=$ $0.907$ $W_{1}$, $a_{2}$ $=$
    $4.372$ $W_{1}$, $T_{0}^{(\IM)}=0.040W_{1}$, $T_{0}^{(\RE)}$ $=$
    $0.038W_{1}$. We emphasize the limited accuracy of
    $T_{0}^{(\IM,\RE)}$ due to the logarithmic nature of the
    fits. Note the contrast to the case $J=0$ (b, d) with the standard
    FL behavior ($\IM\,\Sigma(\omega)\propto-\omega^{2}$ and
    $\RE\,\Sigma(\omega)\propto-\omega$). }
\end{figure}

Both the fully gapped TIAM and the OSMP of the lattice model~\eqref{eq:LattHam}
are thus described by an underscreened spin-$1$ Kondo-type model~\cite{nozieres1980kondo},
which has an intrinsic instability towards ferromagnetism. This quantum
critical behavior is manifest in divergent spin susceptibilities~\cite{PhysRevB.72.014430,PhysRevB.75.245329,PhysRevB.72.045117},
i.e., the density of states for magnetic excitations becomes infinite
at $\omega=0$. In contrast to a standard (local) Fermi liquid, the
metallic properties of such underscreened models are characterized
by a vanishing coherence scale and are referred to as \emph{singular
Fermi liquids} (SFL)~\cite{Varma2002267,PhysRevB.68.220405,PhysRevB.72.045117}.
The self-energy of SFLs is given at low frequencies
by~\cite{wright2011theoretical}
\begin{subequations}%
\begin{align}%
\IM\,\Sigma_{\text{SFL}}(\omega) & =a_{1}\log^{-2}|\omega/T_{0}|+\mathcal{O}(\log^{-4}|\omega/T_{0}|),\label{eq:imSIGMASFL}\\
\RE\,\Sigma_{\text{SFL}}(\omega) & =a_{2}\log^{-3}|\omega/T_{0}|+\mathcal{O}(\log^{-5}|\omega/T_{0}|).\label{eq:reSIGMASFL}
\end{align}%
\end{subequations}%
The scale $T_{0}$ corresponds to the Kondo scale of the underscreened
spin-$1$ impurity, i.e., the energy scale at which the crossover to
the unscreened (residual) spin $1/2$ occurs
\cite{wright2011theoretical,PhysRevB.72.045117}. In
Fig.~\ref{fig:FIG7}a and b
we show fits of Eqs.~\eqref{eq:imSIGMASFL} and \eqref{eq:reSIGMASFL}
to the metallic self-energy $\Sigma_{2}(\omega)$ of the DMFT solution,
confirming the SFL character of the metallic band in the OSMP. By
contrast, the behavior of standard Fermi liquids is observed for
$U_{1}$ $>0$ and $J$ $=$ $0$ (Fig.~\ref{fig:FIG7}b and d).

\emph{Conclusion.---}
Using DMFT, we established that the metallic state in the
OSMP of the two-band Hubbard model is a singular Fermi liquid and
clarified the longstanding question of quantum criticality towards
ferromagnetism of the phase, which was first discussed in the context
of an approximate double-exchange Hamiltonian~\cite{Biermann_PhysRevLett.95.206401}.
We found that a ferromagnetic instability is induced by any nonzero
Hund's rule coupling $J$ $>$ $0$, and since it results from the
effective Kondo physics it will depend only weakly on details such as
the noninteracting band structure. As a consequence, a pure OSMP
ground state of~\eqref{eq:LattHam} is unstable, also in more realistic
multi-band Hubbard models or in correlated materials. However, any
weak interband hybridization is expected to turn the OSMP into a Fermi
liquid with a small coherence
scale~\cite{doi:10.1146/annurev-conmatphys-020911-125045}, which will
lead to an orbital-selective Mott transition at finite temperature, as
observed in the iron pnictide
Rb$_{x}$Fe$_{2-y}$Se$_{2}$~\cite{PhysRevLett.110.067003,2013arXiv1309.6084}.
In any case, the ground state of a system with \emph{selective
  Mottness} \cite{2012arXiv1212.3966D} will be different from the
(unstable) OSMP of the idealized Hamiltonian~\eqref{eq:LattHam}. In
particular, superexchange processes between neighboring lattice sites
can induce antiferromagnetic order, which will compete with the
ferromagnetic instability due to the Hund's rule exchange.

\emph{Acknowledgments.---}
We are grateful to Wilhelm Appelt, Liviu Chioncel, Shintaro Hoshino,
and Dieter Vollhardt for useful dicussions.  This work was supported
in part by the Deutsche Forschungsgemeinschaft through TRR 80.

\bibliographystyle{apsrev4-1}
\bibliography{paper020bib}

\end{document}